\documentclass[aps, prb, a4paper, twoside,twocolumn,fleqn,floatfix]{revtex4}
\usepackage{times,ulem}
\usepackage{graphicx,multirow,amssymb}
\newcommand{\Hmol}{\ensuremath{\widehat{H}_{\text{mol}}}}
\def\hyph{-\penalty0\hskip0pt\relax}
\begin{document}
\normalem
\title{Influence of Functional Groups on Charge Transport in Molecular Junctions}

\author{D. J. Mowbray}
\email{dmowbray@fysik.dtu.dk}
\author{G. Jones}
\author{K. S. Thygesen}
\email{thygesen@fysik.dtu.dk}

\affiliation{Center for Atomic{\hyph}scale Materials Design (CAMD), Department of Physics, Technical
University of Denmark, DK{\hyph}2800 Kgs.\ Lyngby, Denmark}

%
\begin{abstract}
  Using density functional theory (DFT), we analyze the influence of five
  classes of functional groups, as exemplified by NO$_{\text{2}}$, OCH$_{\text{3}}$,
  CH$_{\text{3}}$, CCl$_{\text{3}}$, and I, on the transport properties of a
  1,4{\hyph}benzenedithiolate (BDT) and 1,4{\hyph}benzenediamine (BDA)
  molecular junction with gold electrodes.  Our analysis demonstrates
  how ideas from functional group chemistry may be used to engineer a
  molecule's transport properties, {as was shown experimentally and using a semiempirical model for BDA [Nano Lett.\ \textbf{7}, 502 (2007)]}. 
  In particular, we show that the
  qualitative change in conductance due to a given functional group
  can be predicted from its known electronic effect (whether it is $\sigma$/$\pi$
  donating/withdrawing). However, the influence of functional
  groups on a molecule's conductance is very weak, as was also found in {the BDA}
  experiments. The calculated DFT
  conductances for the BDA species are five times larger than the
  experimental values, but good agreement is obtained after correcting for
  self{\hyph}interaction and image charge effects.
\end{abstract}

\maketitle

\begin{table}
\caption{Functional group categorization by electronic effect with further examples
\cite{OrganicChemistry}.}\label{FunctionalGroups}
\begin{ruledtabular}
\begin{tabular}{ccclc}
\multicolumn{2}{c}{Functional} &Structure& \multicolumn{1}{c}{Electronic} & Other \\
\multicolumn{2}{c}{Group} & & \multicolumn{1}{c}{Effect} & Examples\\\hline
\multirow{2}{*}{NO$_{\text{2}}$}&\multirow{2}{*}{nitro} 
&\multirow{2}{*}{\includegraphics[scale=0.06]{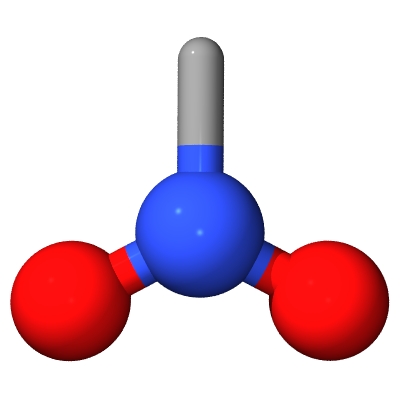}}& \multirow{2}{*}{$\pi$
withdrawal} & CN, COR, \\
&&&&SO$_{\text{3}}$R\\
 \multirow{2}{*}{OCH$_{\text{3}}$}& \multirow{2}{*}{methoxy}  &
\multirow{2}{*}{\includegraphics[scale=0.06]{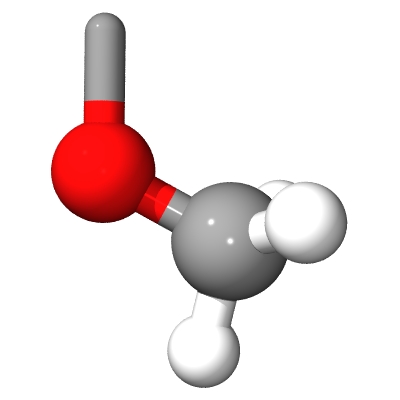}}&\multirow{2}{*}{$\pi$
donation} & \multirow{2}{*}{NR$_{\text{2}}$, OR}\\
&&&&\\
 \multirow{2}{*}{CH$_{\text{3}}$}& \multirow{2}{*}{methyl}  &
\multirow{2}{*}{\includegraphics[scale=0.06]{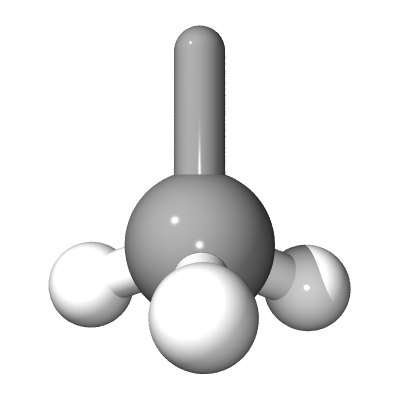}}&\multirow{2}{*}{$\sigma$
donation} & C$_{\text{2}}$H$_5$, C$_{\text{3}}$H$_7$\\
&&&&alkyls\\
\multirow{2}{*}{CCl$_{\text{3}}$}&\multirow{2}{*}{trichloromethyl}  &
\multirow{2}{*}{\includegraphics[scale=0.06]{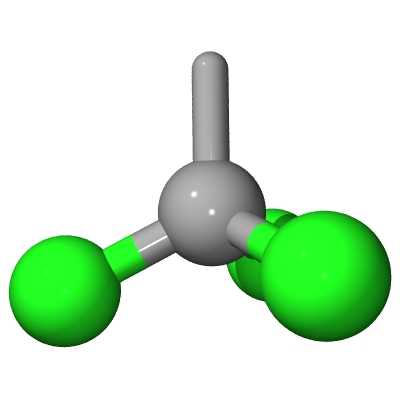}}& \multirow{2}{*}{$\sigma$
withdrawal} & \multirow{2}{*}{CF$_{\text{3}}$, NR$_{3}^+$}\\
&&&\\
 \multirow{2}{*}{I}&\multirow{2}{*}{iodo}  &
\multirow{2}{*}{\includegraphics[scale=0.06]{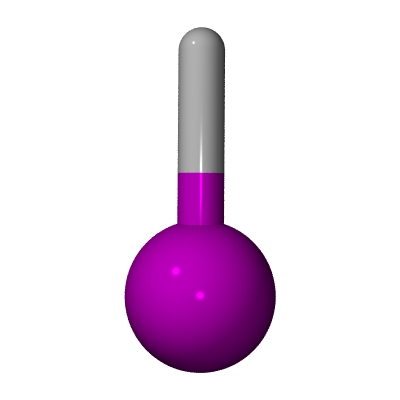}}&$\sigma$ withdrawal \& & F, Cl,
Br\\
&&& $\pi$ donation& halogens
\end{tabular}
\end{ruledtabular}
\end{table}

\begin{figure*}
\begin{center}
\includegraphics[width=6.25in]{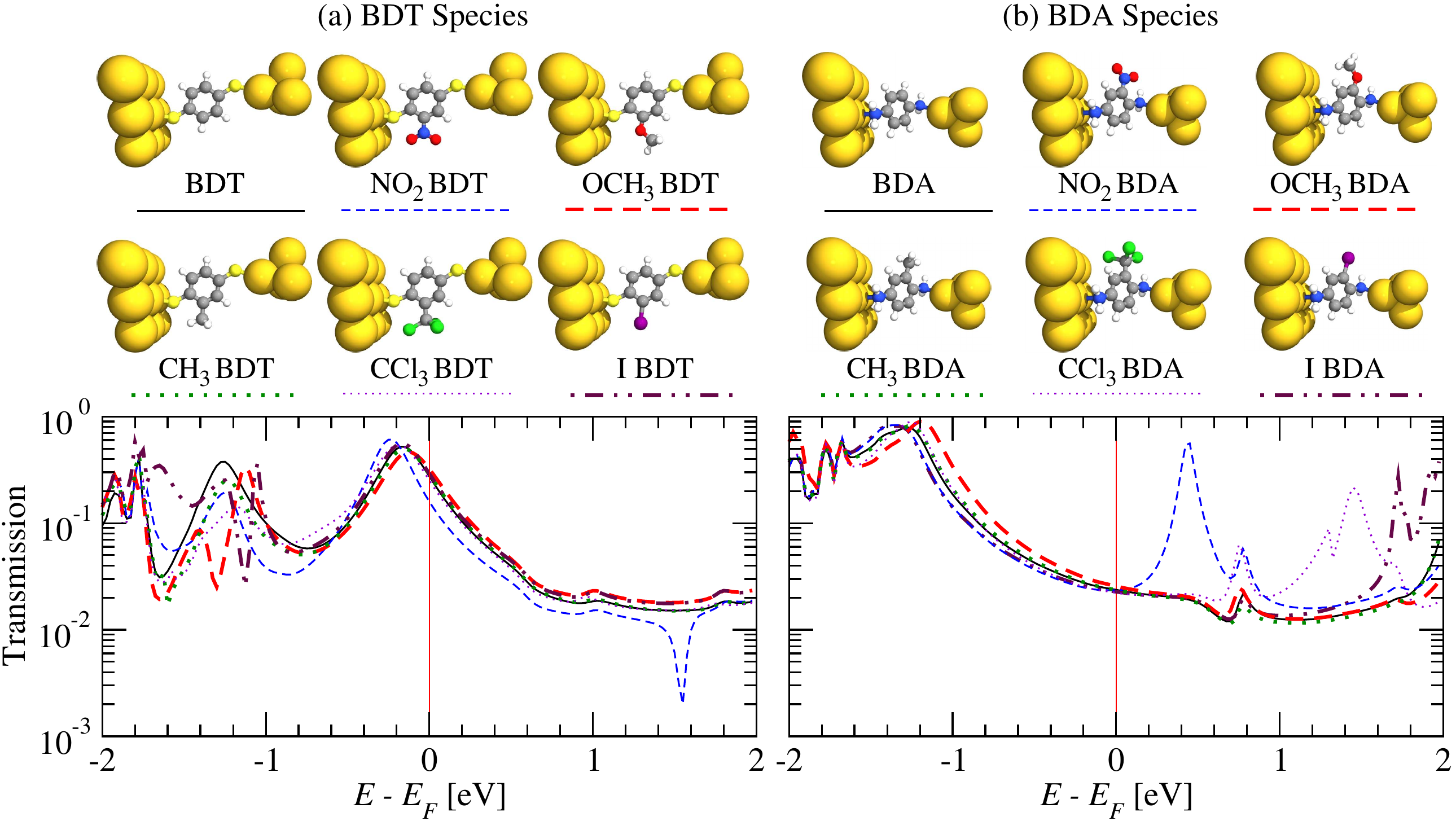}
\end{center}
\caption[]{(Color online). Schematics of (a) BDT species and (b) BDA species bonded
  between a gold (111) surface and tip, and their
  transmission functions versus energy \(E - E_F\) in eV,
  relative to the metal's Fermi energy \(E_F\).  }\label{Transport}
\end{figure*}

One of the main advantages of molecular based electronics over
present day semiconductor technology is the vast flexibility in
design and functionality offered by the myriad of available
molecules~\cite{MonoMolecularElectronics}. In order to fully exploit
this flexibility for the synthesis of molecular devices, it is
vital to establish simple guiding rules to
estimate the effect that a given change in the molecule's composition
or conformation has on its electrical properties.

So far, most experimental and theoretical work on single{\hyph}molecule conduction
has focused on
understanding the basic properties of individual junctions
\cite{Exp_Reed,stokbro,Exp_SC6S_SC8S_SC10S,stadler}. On the other hand, few studies
have aimed at describing general trends across molecular
species~\cite{MolecularConformation,AnchoringGroups,Exp_Ref1}.

It is well known that the chemistry of a
functional parent or parent molecule may be altered in a consistent
manner by the attachment of functional groups~\cite{OrganicChemistry}.
Further, the influence of a particular functional group on a parent
molecule's chemistry may be predicted qualitatively by considering the
electronic effect of the functional group.

We will demonstrate how such ideas may also be applied to describe the
influence of a functional group on a parent molecule's transport
properties. {This was recently shown experimentally for BDA, and the results were explained by scaling the molecule's empirical conductance value by the calculated ionization potentials in the gas phase for the BDA species with two bonded Au atoms~\cite{BDASubstituents}.}  By employing these concepts from functional group
chemistry, we may engineer the electronic transport properties of a
known functional parent.  This may be done using a functional group
whose frontier orbitals have the same symmetry as the conducting orbital
of the parent molecule ($\sigma$ or $\pi$). Groups with donating/withdrawing electronic
effects may then be used to raise/lower the eigenenergy of the conducting orbital.

In this paper, we analyze the influence of functional groups on the
conductance of both a 1,4{\hyph}benzenedithiolate (C$_6$H$_4$S$_{\text{2}}$ or
BDT) and a 1,4{\hyph}benzenediamine (C$_6$H$_4$(NH$_{\text{2}}$)$_{\text{2}}$ or BDA)
molecular junction between gold contacts.  We consider these molecules
because BDT is perhaps the best studied molecule for charge transport
\cite{Exp_Reed,stokbro,Benzenedithiol1},
while the influence of functional groups on the BDA molecule has been
recently studied experimentally~\cite{BDASubstituents}. In
particular, we shall consider prototypical functional groups from each
of the five electronic categories, as shown in Table
\ref{FunctionalGroups}.  These categories are: (1) withdrawal by
conjugation through the \(\pi\) network (\(\pi\) withdrawal), (2)
donation by conjugation through the \(\pi\) network (\(\pi\)
donation), (3) donation by inductive effect through the \(\sigma\)
network (\(\sigma\) donation), (4) withdrawal by inductive effect
through the \(\sigma\) network (\(\sigma\) withdrawal), and (5)
withdrawal by inductive effect through the \(\sigma\) network with
donation by conjugation through the \(\pi\) network (\(\sigma\)
withdrawal \& \(\pi\) donation)~\cite{OrganicChemistry}.  For each
electronic effect categorization we have selected as respective
examples a nitro group (NO$_{\text{2}}$), a methoxy group (OCH$_{\text{3}}$), a methyl
group (CH$_{\text{3}}$), a trichloromethyl group (CCl$_{\text{3}}$), and an iodo group
(I).  Other examples of functional groups from each category are also
provided in Table \ref{FunctionalGroups}.

All calculations have been performed using the SIESTA density
functional theory (DFT) code~\cite{SIESTA} with the PBE
exchange correlation (xc) functional~\cite{PBE}, a double zeta
polarized (DZP) basis set, and a mesh cutoff of 200 Ry.  We first
adsorbed the molecules on a five layer thick gold (111) slab and
relaxed the molecule and the three outermost surface layers. The
supercell contained 3$\times$3 gold atoms in the surface plane, within which we
used 2$\times$2 $k$-points. A pyramid of four gold
atoms attached to the five layers of gold (111) was then introduced in the
supercell to simulate the tip of a scanning tunneling microscope
(STM). 
The pyramid was placed with 
its apex atom close to the unbound
S/N atom of the molecule, and upon further relaxation the
molecule bonds to the gold apex atom, 
as shown in Fig.\ \ref{Transport}.

{In this way we model a low temperature conductance measurement
  for a low molecular coverage on a gold (111) surface using an STM
  tip~\cite{SMCondRev}.  Such a configuration may not accurately
  describe the results of break{\hyph}junction experiments, especially
  for the weakly bound BDA species~\cite{BDA}.  However, {as shown in Ref.~\onlinecite{BDA}, such a binding site yields a similar transmission function to more energetically favorable geometries. Thus, our chosen geometry} should
  allow us to discern more broadly applicable trends in the electronic
  effects of the functional groups, which are our primary interest. } 

We find that the BDT species (BDT, NO$_{\text{2}}$BDT, OCH$_{\text{3}}$BDT,
CH$_{\text{3}}$BDT,
and I BDT) prefer the bridge site of the (111) surface and are rotated by approximately
30\(^\circ\) to the surface normal~\cite{Ref36}, as shown
in Fig.\ \ref{Transport}(a).  
On the other hand, we find that the BDA species (BDA,
NO$_{\text{2}}$BDA, OCH$_{\text{3}}$BDA, CH$_{\text{3}}$BDA, and I BDA) prefer the
atop site
and lie down at an angle of approximately 50\(^\circ\) to the surface normal
\cite{BDA}, as shown in Fig.\
\ref{Transport}(b).  

Having chosen the contact geometries, we calculated the
elastic transmission functions using the non{\hyph}equilibrium Green's
function (NEGF) formalism. The calculation procedure is equivalent
to the one described in Ref.~\onlinecite{Thygesen1}, except
that the Hamiltonian was represented in the SIESTA atomic orbital
basis instead of Wannier functions.  We note here that the DZP SIESTA
basis set has recently been shown to yield transmission functions in
quantitative agreement with plane wave codes and maximally localized
Wannier functions~\cite{BenchmarkPaper}. The transmission function was
averaged over four $k$-points in the surface plane, and three gold (111)
layers were added on both sides of the molecule before the system
was coupled to the bulk leads to ensure a smooth matching of the
effective DFT potential.

The calculated transmission functions are shown in Fig.\
\ref{Transport}. For both BDT and BDA the change in the transmission
function, $T(E)$, near the Fermi level when the functional groups are
attached, is strikingly small.

\begin{table}
\caption[]{Conductance $G$ of BDT and BDA species between a gold (111) surface and
tip.}\label{Conductance}
\begin{ruledtabular}
\begin{tabular}{lclc}
BDT Species & \(G\) [\({\text{2}} e^{\text{2}}/h\)] & BDA Species &\(G\)
[\({\text{2}} e^{\text{2}}/h\)]\\\hline
BDT               &  0.28 & BDA &0.024\\
NO$_{\text{2}}$BDT  & 0.16&NO$_{\text{2}}$BDA &0.024\\
OCH$_{\text{3}}$BDT &0.32&OCH$_{\text{3}}$BDA&0.026\\
CH$_{\text{3}}$BDT  & 0.29&CH$_{\text{3}}$BDA &0.024\\
CCl$_{\text{3}}$BDT &0.25&CCl$_{\text{3}}$BDA &0.022\\
I$_{\ }$BDT        &0.29&I$_{\ }$BDA &0.023\\
\end{tabular}
\end{ruledtabular}
\end{table}

The calculated conductances, \(G = G_0 T(E_F)\), where \(G_0 = {\text{2}}
e^{\text{2}}/h\), for the BDT and BDA species are shown in Table
\ref{Conductance}. As is often the case for DFT transport calculations
on molecular contacts, our results differ substantially from the
experimental values of 0.011 \(G_0\) obtained for BDT~\cite{Benzenedithiol1}, and
0.0064 \(G_0\)
 for BDA~\cite{BDA}. We will address this issue in the last part of the
paper. At this point, we observe that functional groups whose
valence orbitals are of \(\pi\) symmetry seem to have the most influence
on the molecule's conductance, with the \(\pi\) withdrawing group
lowering the conductance, and the \(\pi\) donating group raising the
conductance.

\begin{figure}
\begin{center}
\includegraphics[width=3.1in]{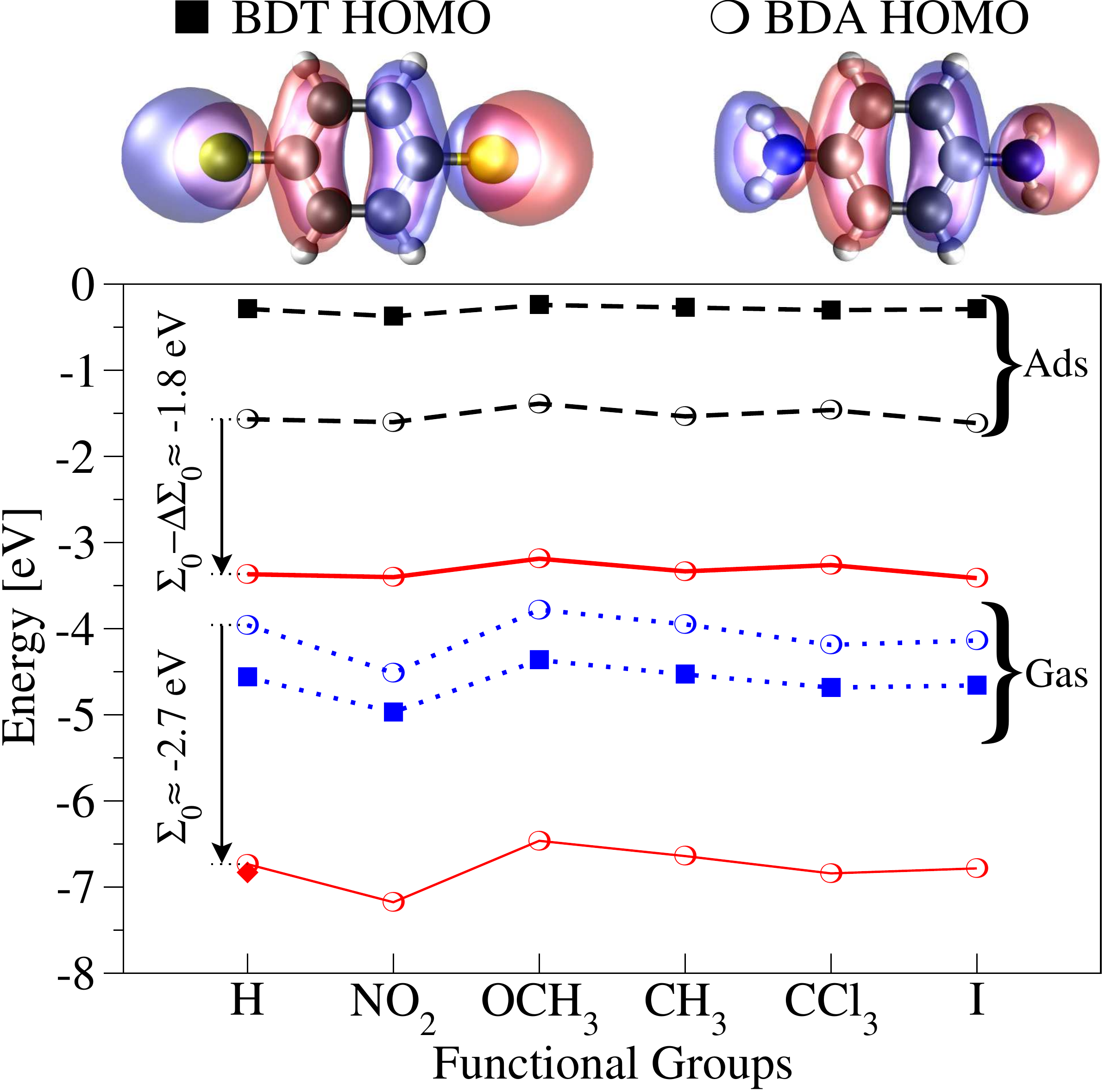}
\end{center}
\caption[]{(Color online). Eigenenergies in eV and isosurfaces of \(\pm\)0.02 \(e/\)\AA\(^{\text
3}\) of HOMOs for BDT species
  (squares) and BDA species (circles), when adsorbed between a gold
  (111) surface and tip relative to the metal's Fermi level $E_{F}$
  (black dashed lines), and in the gas phase relative to 
  vacuum (blue dotted lines). For BDA species,
  the experimental (diamond) and calculated (thin red
  solid line) gas phase ionization potential $I_0$ with a SO shift of
  $\Sigma_0 \approx$ -2.7 eV, and the eigenenergies in the junction with a
  SO shift of \(\Sigma_0 + \Delta\Sigma_0 \approx\) -{1.8} eV (thick red solid line), are
  shown.}\label{Hybrid}
\end{figure}

For both BDT and BDA species, we found that transport primarily occurs
through the HOMO, which in both cases is a \(\pi\) bonding molecular
orbitals, as shown in Fig.\ \ref{Hybrid}. In principle, the concept of
molecular orbitals is somewhat artificial for a chemisorbed molecule,
as hybridization effects will broaden the levels into resonances. One
way of generalizing the concept is to consider $\Hmol$, the projection of the
Hamiltonian of the \textit{contacted} system onto the subspace spanned
by the basis functions of the molecule. The
eigenvectors of $\Hmol$ can be regarded as
molecular orbitals renormalized by the electrodes~\cite{Thygesen1}. For molecules which are not too strongly
coupled to the electrodes, these renormalized molecular levels are
easily identified with the levels of the free molecule.

In Fig.\ \ref{Hybrid} we compare the position of the
HOMO level (relative to $E_F$) of the contacted molecules with the HOMO
level (relative to vacuum) of the free gas phase molecules.
The variation in the HOMO position with the functional group
correlates directly with the variation in the conductance (see Table
\ref{Conductance}) which shows that the current is indeed carried by
the HOMO. One exception from this trend is NO$_{\text{2}}$BDA for which the conductance is not lowered even though the HOMO is significantly down shifted. This is because the LUMO of NO$_{\text{2}}$BDA lies so low that it also contributes to the charge transport. From Fig.\ \ref{Hybrid} we can also see that each functional group
produces a remarkably consistent
shift of the conducting orbital for BDA and BDT. This agrees
with the basic premise of functional group chemistry, that a given
functional group will alter different parent molecules in a consistent
way. We also notice that the
two \(\pi\) groups (NO$_{\text{2}}$ and OCH$_{\text{3}}$) give the largest shifts of
the
eigenenergies, while the functional groups with \(\sigma\) symmetry
have little influence. Moreover, the $\pi$ withdrawing functional
group ($\text{NO}_{\text{2}}$)
delocalizes the HOMO and thereby lowers the
eigenenergy, while the $\pi$ donating functional group ($\text{OCH}_{\text{3}}$)
confines the HOMO and thereby increases the eigenenergy.
Thus the
main ideas of functional group chemistry may be employed to
qualitatively predict a functional group's influence on a given
functional parent's conducting orbital(s).

We stress that the qualitative effect of the functional groups on the HOMO of the
contacted molecule, and thus its conductance, roughly follows from the effect of the
functional groups on the free molecule's HOMO{, as was found experimentally for BDA~\cite{BDASubstituents}}. This is significant, as it suggests that the latter could be used as a simple descriptor to estimate the impact of a given functional group on the conductance, thereby allowing for an efficient screening of large numbers of functional groups.

When we compare in Fig.\ \ref{Hybrid} the magnitude of the
shifts due to the functional groups for both BDT and BDA,
we find significantly smaller effects for the contacted molecules than for the gas phase molecules.  
For example, the shift
due to NO$_{\text{2}}$ on BDT is approximately 0.4 eV in the gas phase,
while it is less than 0.1 eV in the contact.
This may be understood by recognizing that the gold contacts act as electron
sources/sinks for the parent molecule, counteracting the functional group's
electronic influence. We thus believe that the weak influence of the functional groups on the conductance is a result of the self{\hyph}consistent `pinning' of the HOMO level which ensures the charge
neutrality of the molecule. Such effects are indeed physical, but
could be artificially enhanced by self{\hyph}interaction errors in the PBE xc{\hyph}functional: For partially occupied molecular orbitals, an incomplete cancellation
of the Coulomb self interaction by the xc{\hyph}functional will
artificially raise (lower) the energy of that orbital when charge is
added (removed) and this will enhance the `level pinning'.

Self interaction errors also contribute to the
well{\hyph}known underestimation of band gaps by DFT calculations
\cite{BandGapErrors}.  Recently, an atomic self
interaction correction (ASIC) scheme~\cite{SICTheory,ASIC} has been
proposed as a simple cure to this problem for molecular contacts. In
general, however, image charges formed in the metallic electrodes when
electrons are added to or removed from the molecule also renormalize
the molecular levels~\cite{LevelRenomalization}, and this effect is not
captured by the SIC.

\begin{table}
\caption[]{(Color online). Calculated conductance $G_{\text{calc}}$, SO conductance
$G_{\text{SO}}$~\cite{BDA} with shift \(\Sigma_0 - \Delta\Sigma_0 \approx \)-{1.8} eV, and experimental conductance $G_{\text{exp}}$~\cite{BDASubstituents} for
BDA species between a gold (111) surface and tip.}\label{PSIC_exp}
\begin{ruledtabular}
\begin{tabular}{lccc}
 BDA Species  &\(G_{\text{calc}}\) [\({\text 2}e^{\text{2}}/h\)]&\(G_{\text{SO}}\) 
[\({\text 2}e^{\text{2}}/h\)]&
\(G_{\text{exp}}\) [\({\text 2}e^{\text{2}}/h\)]\\\hline
 BDA               &0.024 &{0.0035}&0.0064\footnotemark[1]\footnotetext[1]{Ref.~\onlinecite{BDASubstituents}}\\
NO$_{\text{2}}$BDA   &0.024 &{0.0032}& ---\\
OCH$_{\text{3}}$BDA  &0.026 &{0.0035}&{0.0069}\footnotemark[1]\\
CH$_{\text{3}}$BDA   &0.024 &{0.0034}&{0.0064}\footnotemark[1]\\
CCl$_{\text{3}}$BDA  &0.022 &{0.0034}& ---\\
Cl$_{\ }$BDA        &0.015 &{0.0025}&{0.0060}\footnotemark[1]\\
I$_{\ }$BDA         &0.023 &{0.0036}& ---\\
\end{tabular}
\end{ruledtabular}
\end{table}

In Table \ref{PSIC_exp} we compare our DFT calculated conductances for
the BDA species with corresponding experimental
values~\cite{BDASubstituents}. The calculated numbers are roughly five
times larger than the experimental ones due to the above mentioned
deficiencies of the DFT approach. It is, however, interesting to
notice that: (i) the weak effect of functional groups is found in both
data sets (ii) the qualitative effect of the functional groups on the
conductance (increase/decrease) is reproduced by calculations. In the
second column of Table \ref{PSIC_exp} we present the results of our
conductance calculations after a `scissors operator' (SO), which corrects
for self{\hyph}interaction errors and image charge effects, has been applied
to the molecule's spectrum. 

To construct the SO we follow a recent work by Quek \textit{et al.}\ who applied the scheme to a BDA{\hyph}gold junction~\cite{BDA}.  
In this method, the underestimation of the DFT HOMO {and LUMO} are corrected by
shifting the occupied{/unoccupied} orbitals of the contacted molecule by
\({\Sigma_0^{o,u} = \pm}\Sigma_0 = {\mp(I_0 + \varepsilon_{\text{HOMO}})}\), where
\(\varepsilon_{\text{HOMO}}\) is the DFT HOMO level for the \textit{free} molecule and \(I_0\) the ionization potential {\cite{LUMONote}}. We also calculate the
later in DFT from \(I_0 = E_{q=+{\text{1}}} - E_{q={\text{0}}}\), where
\(E_{q=+{\text{1}}}\) is the total energy of the molecule with charge \(+e\) in
the gas phase, while \(E_{q={\text{0}}}\) is the total energy in the gas phase
of the neutral molecule.  For BDA without functional groups we find
\(I_0 \approx\) 6.73 eV, which agrees
quantitatively with the experimental value of 6.83 eV
\cite{BDA}. We found only small variations in \(\Sigma_0\) for the
different BDA species and have used \(\Sigma_0 \approx \) -2.7 eV for all
molecules, as indicated in Fig.\ \ref{Hybrid}.

To estimate the shift of the molecular levels by image
charge effects, we first calculate
the charge distribution after removing an electron from the free
molecule. A Mulliken analysis is then used to approximate this continuous charge
distribution by point
charges located at the atoms of the molecule.  We model the gold contacts as two
perfectly conducting surfaces separated by 13.2 \AA, which corresponds to the vacuum
separation between the opposing gold (111) surfaces. We then obtain an image
potential \(\Delta\Sigma_0 \approx \) -{0.9} eV for all of the BDA species considered
\cite{supplementarymaterial}.  For the BDA
molecular junctions, we thus shift all occupied{/unoccupied} orbitals (obtained by
diagonalizing $\Hmol$) by \({\Sigma_0^{o,u} -
\Delta\Sigma_0^{o,u} = \pm(}\Sigma_0 - \Delta\Sigma_0{)}\approx {\mp}\){1.8} eV as indicated in Fig.\
\ref{Hybrid}. Calculating the conductance using the renormalized
Hamiltonian yields the values shown in Table \ref{PSIC_exp}.  
We find that this `ad hoc' correction produces results {approximately half those obtained from break{\hyph}junction experiments on the BDA species~\cite{BDASubstituents}. This is most probably due to our choice of contact geometry, as our calculated conductance for BDA of 0.024 \(G_0\) is about half the average DFT value calculated over 15 different break{\hyph}junction contact geometries of 0.046 \(G_0\) ~\cite{BDA}. }
We stress, however, that the use of this 
SO can only be justified for weakly coupled molecules with
HOMO/LUMO levels well separated from $E_F$.  In Fig.\ \ref{Hybrid} we
see this is the case for the BDA species, since the HOMOs are
localized near the amine contact groups, providing a poor overlap with
the gold contact orbitals.  However, for the BDT species, whose HOMOs
are rather diffuse around the sulfur atoms, there is a strong overlap
with the gold contact orbitals, so that application of the SO is not justified.

In conclusion, we found that the ideas of functional group chemistry may be applied to
qualitatively predict the influence of a functional group on the
electronic structure of a parent molecule{, as has been found for BDA both experimentally and using a semiempirical model~\cite{BDASubstituents}}.
However, we also found that functional groups have a very weak
influence on a molecule's conductance in agreement with recent experiments~\cite{BDASubstituents}. The reason for the weak influence is that charge neutrality pins the HOMO/LUMO molecular levels, making it difficult to shift them relative to $E_F$.
By applying multiple functional groups to the same parent
molecule it may be possible to obtain a stronger influence, as was found for BDA
\cite{BDASubstituents}. By employing a scissors operator correcting for self{\hyph}interaction errors and image charge effects, we obtained {qualitative} agreement with experimental conductance values for the BDA species.  
Our results suggest that effective `switching' of a molecule's
conductance may require a more direct change in the strength or
geometry of the molecule's contacts, to overcome the `level pinning'
of the metal.

We thank J. K. N{\o}rskov, S. Dobrin, and M. Strange for useful
discussions. The authors acknowledge support from the Danish Center for
Scientific Computing through grant No. HDW-1103-06. The Center for Atomic{\hyph}scale
Materials Design (CAMD) is sponsored by the
Lundbeck Foundation.

\appendix

\section{Scissors Operator (SO)}\label{Appendix 1}

The SO for a free molecule in the gas phase \(\Sigma_0\), is given by
\begin{eqnarray}
\Sigma_0 \!\!&=&\!\! - (\varepsilon_\mathrm{HOMO} + I_0),
\end{eqnarray}
where \(\varepsilon_\mathrm{HOMO}\) is the DFT HOMO level of the free molecule in the gas phase, and \(I_0\) is its first ionization potential, which is defined as
\begin{eqnarray}
I_0 \!\!&=&\!\! E_{q=+1} - E_{q=0}.
\end{eqnarray}
Here \(E_{q=+1}\) is the total gas phase energy calculated for a molecule with charge \(q = +e\), while \(E_{q=0}\) is the total energy in the gas phase for the neutral molecule.

The occupied/unoccupied orbitals of the contacted molecule are shifted by the same amount \(\Sigma_0^{o,u} = \pm \Sigma_0\) for each orbital.  This is reasonable for orbitals other than the HOMO and LUMO, since these orbitals are sufficiently far from the Fermi level \(E_F\) to have little influence on the contacted molecule's conductance.  

However, this is not the case for the LUMO orbital, especially for NO$_2$ BDA.  In this case, a fully rigorous calculation would shift the DFT LUMO level by \(-(\varepsilon_{\mathrm{LUMO}} + E_{ea})\), where \(\varepsilon_{\mathrm{LUMO}}\) is the DFT LUMO level for the free molecule and \(E_{ea}\) the electron affinity.  The later may be calculated similarly to the ionization potential in DFT from \(E_{ea} = E_{q=0} - E_{q=-1}\), where \(E_{q=-1}\) is the total energy of the molecule with charge \(-e\) in the gas phase.  However, as discussed in Ref. \onlinecite{BDA}, when such a shift was calculated for the BDA LUMO, it was found to be approximately equal and opposite to that for the HOMO.  For these reasons, we use \(\Sigma_0^{o,u} = \pm \Sigma_0\) for all the occupied/unoccupied orbitals.

For a molecule adsorbed near a metal surface, the SO is \(\Sigma_0 - \Delta\Sigma_0\), where \(\Delta\Sigma_0\) is the charged molecule's image potential arising from the metal surface.  To estimate the charge distribution on the molecule \(\rho({\bf r})\), we employ the gas phase Mulliken analysis from a charged spin polarized calculation, and use the distribution of unpaired spin up charges.  Table \ref{BDA_Q}(a) and \ref{BDA_Q}(b) show the distribution of charges on each atom in (a) 1,4\hyph benzenediamine (BDA) and (b) 3\hyph nitro\hyph 1,4\hyph benzenediamine (NO$_2$BDA) respectively.  
\begin{table}[!b]
\caption{Mulliken analysis of charge distribution on (a) BDA and (b) NO$_2$BDA in the gas phase with total charge \(q = +e\).}\label{BDA_Q}
\begin{ruledtabular}
\begin{tabular}{lcrrr}
(a) 1,4{\hyph}benzenediamine &\(Q_i\) [e] & \(x_i\) [\AA] & \(y_i\) [\AA] & \(z_i\) [\AA]\\\hline
\multirow{12}{*}{\includegraphics[height=0.5\columnwidth]{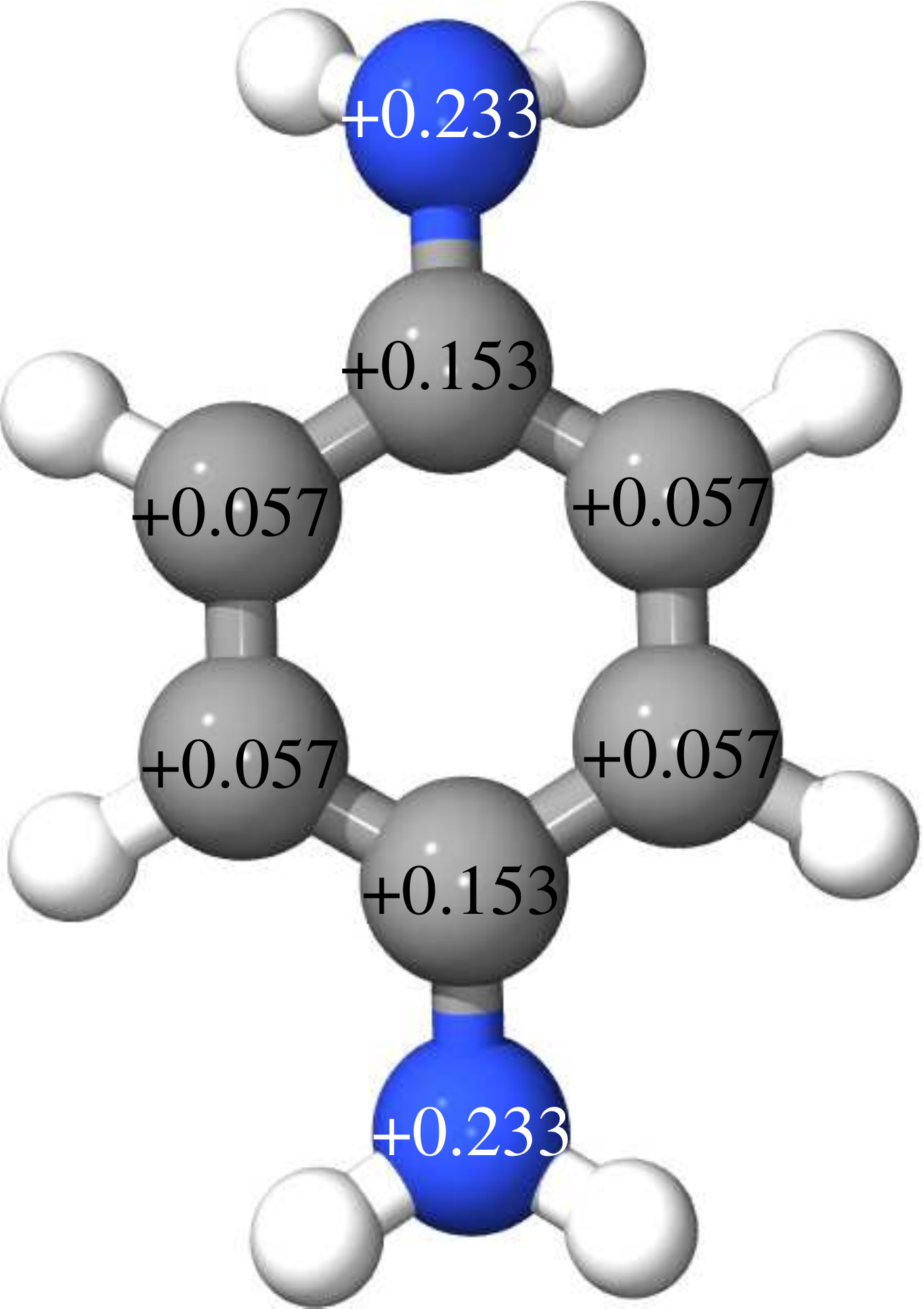}}\\
&+0.233 & 2.508 & 2.532 & 4.225\\
&+0.153 & 3.435 & 2.297 & 5.257\\
&+0.057 & 3.770 & 0.972 & 5.657\\
&+0.057 & 4.027 & 3.390 & 5.954\\
&+0.057 & 4.766 & 0.754 & 6.635\\
&+0.057 & 5.028 & 3.165 & 6.916\\
&+0.153 & 5.438 & 1.846 & 7.243\\
&+0.233 & 6.553 & 1.663 & 8.141\\
& \\
\hline\hline
(b) 3{\hyph}nitro,1,4{\hyph}benzenediamine &\(Q_i\) [e] & \(x_i\) [\AA] & \(y_i\) [\AA] & \(z_i\) [\AA]\\\hline
\multirow{12}{*}{\includegraphics[height=0.5\columnwidth]{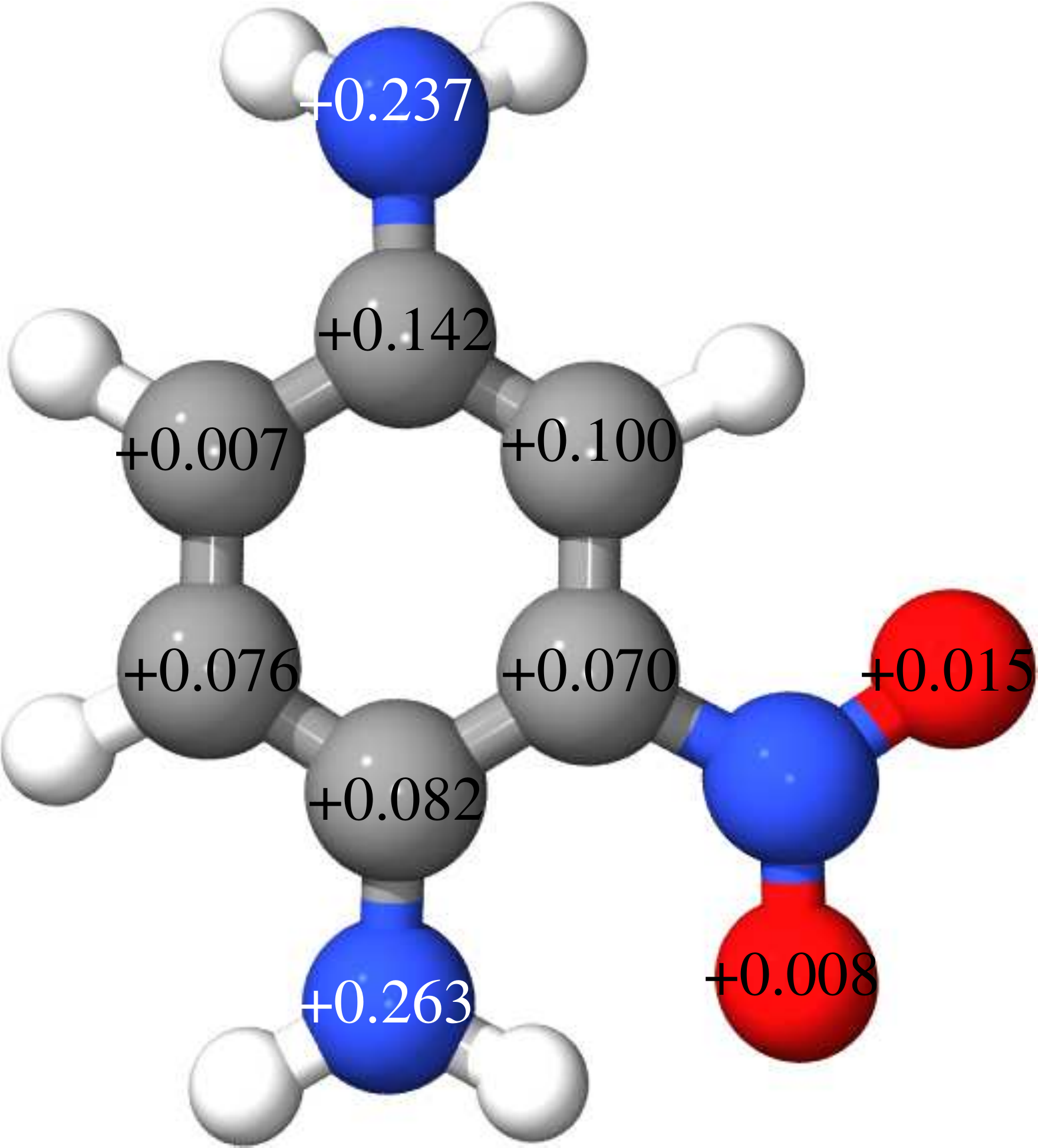}}
&+0.237 & 2.591 & 2.540 & 4.187\\
&+0.142 & 3.504 & 2.302 & 5.220\\
&+0.007 & 4.089 & 3.397 & 5.932\\
&+0.100 & 3.834 & 0.989 & 5.626\\
&+0.076 & 5.071 & 3.177 & 6.909\\
&+0.070 & 4.822 & 0.773 & 6.626\\
&+0.015 & 4.361 &-1.509 & 6.479\\
&+0.082 & 5.506 & 1.865 & 7.258\\
&+0.263 & 6.601 & 1.694 & 8.149\\
&+0.008 & 5.979 &-0.878 & 7.847\\
\end{tabular}
\end{ruledtabular}
\end{table}

The total potential energy gain by moving the charged molecule to the metal surface is
\begin{eqnarray}
\Delta\Sigma_0 \!\!&=&\!\! \!\!\int\!\!d^3{\bf r} \rho({\bf r})\Phi_\mathrm{ind}({\bf r}) + \frac{1}{2} \!\!\int\!\!d^3{\bf r}\rho_\mathrm{ind}({\bf r})\Phi_\mathrm{ind}({\bf r}),
\end{eqnarray}
where the first term is the self energy of the molecule's charge distribution \(\rho({\bf r})\) due to the electrostatic potential \(\Phi_\mathrm{ind}({\bf r})\) induced on the metal surface, 
\begin{eqnarray}
\Phi_\mathrm{ind}({\bf r}) \!\!&=&\!\! \int\!\!d^3{\bf r}'\frac{\rho_\mathrm{ind}({\bf r}')}{\|{\bf r} - {\bf r}'\|},
\end{eqnarray}
and the second term is the Hartree energy for the charge distribution \(\rho_\mathrm{ind}({\bf r})\) induced on the metal surface.   Approximating our gold surface by a perfect conductor, we find the induced charge distribution is restricted to the surface, \(\rho_\mathrm{ind}({\bf r}) = \sigma({\bf r}^\perp)\delta(x)\), where \(\sigma({\bf r}^\perp)\) is the induced surface charge distribution, and \({\bf r}^\perp = y{\bf e}_y + z{\bf e}_z\) are the components of \({\bf r}\) in the \(yz\){\hyph}plane, so that
\begin{eqnarray}
\Delta\Sigma_0 \!\!&=&\!\! \int\!\!d^3{\bf r} \rho({\bf r})\Phi_\mathrm{ind}({\bf r})\ + \left.\frac{1}{2} \!\!\int\!\! d^2{\bf r}^\perp \sigma({\bf r}^\perp)\Phi_\mathrm{ind}({\bf r}^\perp)\right|_{x=0}\!\!\!\!\!\!\!\!\!.
\end{eqnarray}
Using the boundary condition that the total electrostatic potential is zero on the metal surface, we find \(\Phi_\mathrm{ind}({\bf r}^\perp)|_{x=0} = - \Phi_\mathrm{mol}({\bf r}^\perp)|_{x=0}\), where \(\Phi_\mathrm{mol}({\bf r})\) is the Coulomb potential due to the charged molecule.  We then find
\begin{eqnarray}
\Delta\Sigma_0 \!\!&=&\!\! \int\!\!d^3{\bf r} \rho({\bf r})\Phi_\mathrm{ind}({\bf r}) - \left.\frac{1}{2} \!\!\int\!\!d^2{\bf r}^\perp \sigma({\bf r}^\perp)\Phi_\mathrm{mol}({\bf r}^\perp)\right|_{x=0}\!\!\!\!\!\!\!\!\!,\\
&=& \!\!\int\!\!d^3{\bf r} \rho({\bf r})\Phi_\mathrm{ind}({\bf r}) - \frac{1}{2} \!\!\int\!\!d^3{\bf r}\rho_\mathrm{ind}({\bf r})\Phi_\mathrm{mol}({\bf r}),\\
&=& \frac{1}{2} \!\!\int\!\!\!\int\!\! d^3{\bf r}d^3{\bf r}' \frac{\rho({\bf r})\rho_\mathrm{ind}({\bf r'})}{\|{\bf r} - {\bf r}'\|}.
\end{eqnarray}
\begin{figure}[!b]
\includegraphics[width=\columnwidth]{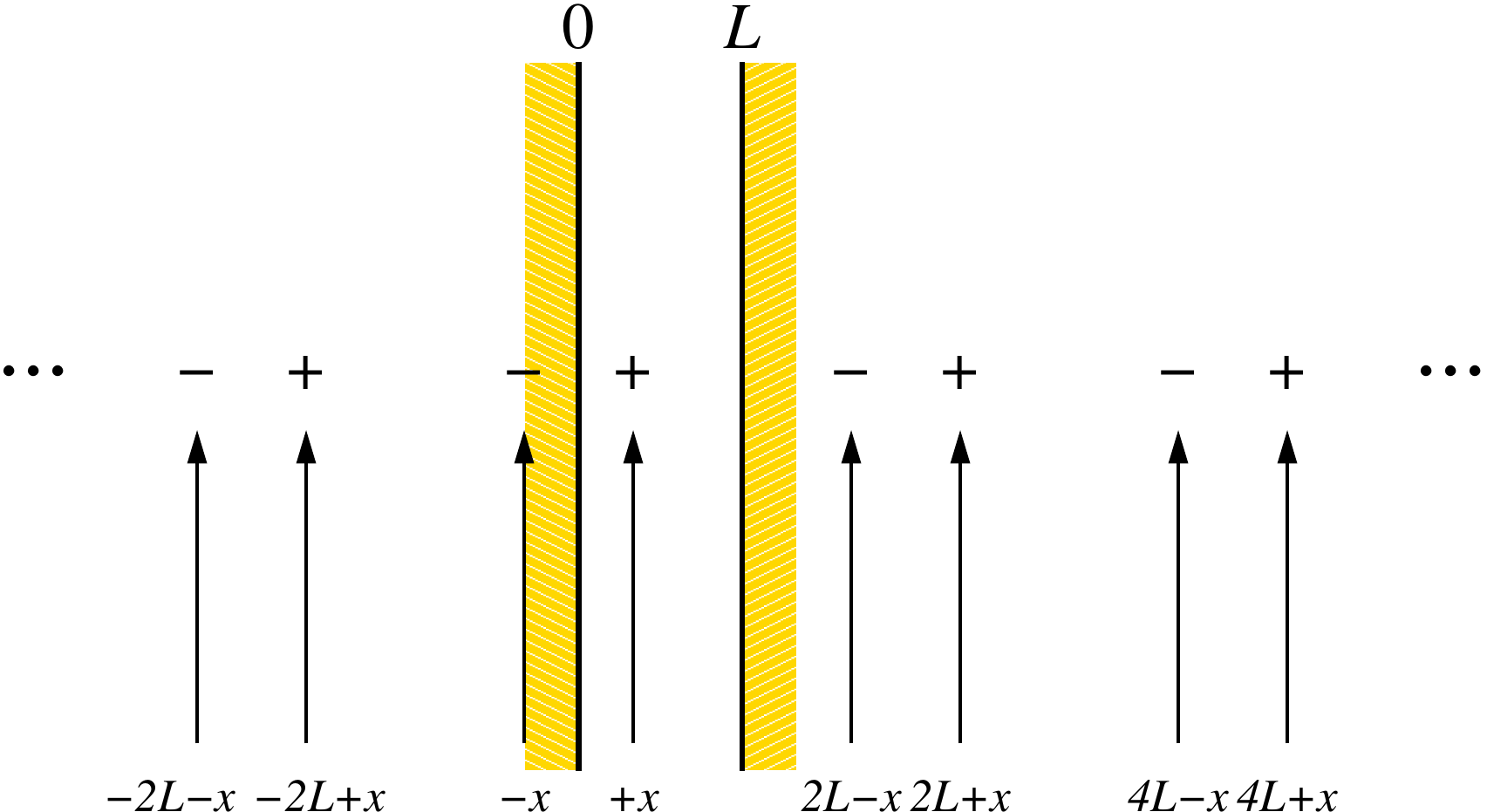}
\caption{Locations of image charges for two perfectly conducting surfaces at $x = 0$ and $x=L$ due to a positive charge at \(+x\), where \(L = \) 13.2 \AA\ for our system.}\label{imagecharges}
\end{figure}

Since we are modeling the molecule's charge distribution as a series of \(N\) point charges at locations \({\bf r}_i = \{x_i,y_i,z_i\}\) with charges \(Q_i\), so that
\begin{eqnarray}
\rho({\bf r}) = \sum_{i=1}^N Q_i \delta({\bf r}-{\bf r_i}),
\end{eqnarray}
and we will only be evaluating the induced potential outside the metal, we may employ the method of image charges, so that
\begin{eqnarray}
\Delta\Sigma_0 \!\!&=&\!\! \frac{1}{2}\!\!\int\!\!\!\int\!\!d^3{\bf r}d^3{\bf r}' \frac{\rho({\bf r})\rho_{\mathrm{img}}({\bf r}')}{\|{\bf r} - {\bf r}'\|},\label{DS0}
\end{eqnarray}
where we have replaced the induced charge density by the image charge density \(\rho_{\mathrm{ind}}({\bf r})\) in (\ref{DS0}).  We may then write
\begin{eqnarray}
\rho_\mathrm{img}({\bf r}) \!\!&=&\!\! \sum_{j=1}^N Q_j \rho_\mathrm{img}^{(j)}({\bf r}),
\end{eqnarray}
so that
\begin{eqnarray}
\Delta\Sigma_0 \!\!&=&\!\! \frac{1}{2}\sum_{i,j=1}^N Q_i Q_j \!\!\int\!\! d^3{\bf r} \frac{\rho_\mathrm{img}^{(j)}({\bf r})}{\|{\bf r_i} - {\bf r}\|},
\end{eqnarray}
where \(\rho_\mathrm{img}^{(j)}({\bf r})\) is the image charge density in the metal due to a point charge of 1 $e$ located at ${\bf r}_j$.

If the metal our molecule is adsorbed upon consists of a single perfect conducting surface located at \(x = 0\), 
we may model the charge distribution induced in the metal as a series of image charges, so that
\begin{eqnarray}
\rho_\mathrm{img}^{(j)}({\bf r}) \!\!&=&\!\! - \delta(x + x_j)\delta({\bf r}^\perp - {\bf r}_j^\perp),\\
\Delta\Sigma_0 \!\!&=&\!\! - \frac{1}{2}\sum_{i,j=1}^N \frac{Q_i Q_j}{\sqrt{(x_i + x_j)^2 + R_{ij}^2}},
\end{eqnarray}
where \(R_{ij} = \|{\bf r}_i^\perp - {\bf r}_j^\perp\|\).

However, if we now introduce a second surface located at \(x = L\), an infinite series of image charges is produced, analogous to a series of repeated reflections between two mirrors.  We then find
\begin{widetext}
\begin{eqnarray}
\rho_\mathrm{img}^{(j)}({\bf r}) \!\!&=&\!\! -\delta({\bf r}^\perp - {\bf r}_j^\perp)\sum_{n=1}^\infty \left[\delta(x + x_j + 2(n-1)L)\right.  + \delta(x + x_j - 2nL) + \delta(x - x_j + 2nL) \left. + \delta(x + x_j - 2nL)\right],\\
\Delta\Sigma_0 \!\!&=&\!\! - \frac{1}{2}\sum_{i,j=1}^N Q_i Q_j \sum_{n=1}^\infty \left[ \frac{1}{\sqrt{(x_i + x_j - 2nL)^2 + R_{ij}^2}}\right.  + \frac{1}{\sqrt{(x_i + x_j + 2(n-1)L)^2 + R_{ij}^2}}\nonumber\\ &&  - \frac{1}{\sqrt{(x_i - x_j + 2nL)^2 + R_{ij}^2}}   \left.- \frac{1}{\sqrt{(x_i - x_j + 2nL)^2 + R_{ij}^2}}\right],
\end{eqnarray}
\end{widetext}
as shown in Fig.\ \ref{imagecharges}.  We shall take the separation between the opposing gold (111) surfaces of 13.2 \AA\ as our \(L\) value. It should be noted that the image charges produced in such a manner are not physical charges, but are employed as mathematical tools to ensure that the total electrostatic potential is zero on both perfect conducting surfaces.

\end{document}